# PLASMA KINETICS IN ELECTRICAL DISCHARGE IN MIXTURE OF AIR, WATER AND ETHANOL VAPORS FOR HYDROGEN ENRICHED SYNGAS PRODUCTION

*A.I. Shchedrin[1], D.S. Levko[1], A.V. Ryabtsev[1], V.Ya. Chernyak[2], V.V. Yukhymenko[2], S.V. Olszewski[2], V.V. Naumov[2], I.V. Prysiazhnevych[2], E.V. Solomenko[2], V.P. Demchina[3], V.S. Kudryavtsev[3]*

[1]*Institute of Physics NANU, Pr. Naiki 46, 03028, Kiev, Ukraine,* [2]*Radiophysical Faculty Taras Shevchenko Kiev National University, Pr. Acad. Glushkova 2/5, 03122, Kiev, Ukraine,* [3]*Gas Institute NANU, Degtyarevskaya St. 39, 03113, Kiev, Ukraine*
*e-mail: aShched@iop.kiev.ua*

**Abstract –** The complex theoretical and experimental investigation of plasma kinetics of the electric discharge in the mixture of air and ethanol-water vapors is carried out. The discharge was burning in the cavity, formed by air jets pumping between electrodes, placed in aqueous ethanol solution. It is found out that the hydrogen yield from the discharge is maximal in the case when ethanol and water in the solution are in equal amounts. It is shown that the hydrogen production increases with the discharge power and reaches the saturation at high value. The concentrations of the main stable gas-phase components, measured experimentally and calculated numerically, agree well in the most cases.

## 1. Introduction

Now there is a great interest in searching of biofuels as an alternative to traditional fossil fuels and natural gas. Bio-ethanol can be a good candidate since it can be obtained in sufficient amounts from agricultural biomass. However, pure ethanol (ethyl alcohol $C_2H_5OH$) has a set of physico-chemical limitations including a relatively low heat of combustion and low speed of ignition. As is known an addition of light and easily burning components ($H_2$, CO, etc) to heavy hydrocarbons significantly increases their combustibility. One possible way is to use plasma reforming of ethanol into hydrogen-enriched synthesis gas (syngas) [1]. The presence of $H_2$ in the mixture allows using ethanol as a real fuel since the speed of flame in hydrogen is ten times higher than in ethanol.

The present paper is related to the study of a new method of the ethanol enrichment by hydrogen using non-equilibrium plasma of a gas discharge in aqueous ethanol solution [2]. In such plasma the energy of neutral particles is much less than the energy of electrons initiating chemical transformations. Selecting the reactions which products are more stabile to the electron impact than initial reagents, one can get a process which is selective to the desired products. In such plasma-liquid system there is no need of removal of the excess energy of the thermal motion of gas particles as they are really cold in plasma.

## 2. Experimental conditions and model of discharge

The schematic of the experimental reactor which was utilized for the plasma-chemical conversion of ethanol into hydrogen is shown in Fig. 1. The advantages of this design are efficiency, compactness and easiness in operation. Two hollow tubes with inserted rod electrodes were placed in the reactor filled up by the ethanol-water solution. The atmospheric air was pumped through the tubes in the gap between the electrodes. The water and ethanol evaporated in the appearing cavity, and the gas discharge burned in the mixture of air and ethanol-water vapors. In experiments, the discharge worked in the continuous regime, typical discharge power was 100 W; the air flow rate was 38 $cm^3/s$, the processing time varied within 1-10 min. The plasma conditions in the discharge were diagnosed by the optical emission spectroscopy; the output syngas products after the reactor were analyzed by the mass-spectrometry and gas-chromatography [2].

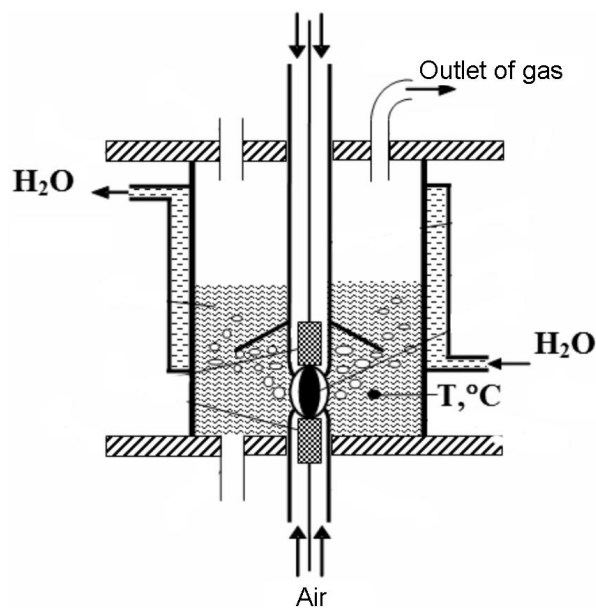

*Fig. 1. Schematic of experimental set-up*

In the model statement it was supposed that the gas discharge was burning in a cylinder cavity with a radius that equals to the internal radius of the tubes and with a length that equals to the distance between the electrodes. In calculations, the complete time of the discharge burning was divided into the equal time intervals which duration is determined by the cavity filling time. In the given case this time is determined only by the time of the gas flowing that is equal to the ratio of the cavity volume $V$ to the gas flow rate $G$, i.e. $\tau = V/G \approx 10^{-3}$ s. In addition it was assumed that at the beginning of the every time interval the gas in the cavity was totally refreshed, and the previous periods did not influence on the subsequent periods. This allows doing calculations of plasma-chemical kinetics in the discharge during the one time interval only as the concentrations of components in the every time interval come to the same values. The gas products from the discharge cavity entered the solution volume in the reactor and then passed into the chamber where the gas composition measurements took place. Here the plasma-chemical kinetics was also calculated but without electron-molecular interactions.

After the detailed analysis of plasma-chemical reactions in the air-water-ethanol mixture, 59 components were taken into account for calculation, and the following system of kinetic equations was used [3]:

$$\frac{dN_i}{dt} = S_{ei} + \sum_j k_{ij} N_j + \sum_{m,l} k_{iml} N_m N_l + \ldots,$$

where $N_i$, $N_j$, $N_m$, $N_l$ are concentrations of molecules, atoms and radicals, $k_{ij}$, $k_{iml}$ are rate constants of chemical reactions for corresponding reagents. The rates of formation of products of electron-molecular reactions $S_{ei}$ were determined by equations:

$$S_{ei} = \frac{W}{V} \frac{1}{\varepsilon_{ei}} \frac{W_{ei}}{\sum_i W_{ei} + \sum_i W_i}.$$

where $W$ is a discharge power, $V$ is a discharge cavity volume, $W_{ei}$ is a specific power consumed in the electron-molecular process of inelastic scattering with threshold energy $\varepsilon_{ei}$:

$$W_{ei} = \sqrt{\frac{2q}{m}} n_e N_i \varepsilon_{ei} \int_0^\infty \varepsilon Q_{ei}(\varepsilon) f(\varepsilon) d\varepsilon,$$

where $q = 1.602 \cdot 10^{-12}$ erg/eV, $m$ and $n_e$ are the mass and concentration of electrons; $Q_{ei}$ is a cross-section of the corresponding inelastic process; $f(\varepsilon)$ is the electron energy distribution function (EEDF). $W_i$ is a specific power spent into the gas heating:

$$W_i = \frac{2m}{M_i} \sqrt{\frac{2q}{m}} n_e N_i \int_0^\infty \varepsilon^2 Q_i(\varepsilon) f(\varepsilon) d\varepsilon,$$

where $M_i$ is a molecular mass, $Q_i$ is a transport cross-section of elastic scattering.

The EEDF was calculated from the Boltzmann equation in the standard two-term approximation [4]. It was assumed that the electric field in the discharge did not vary ($E = 20$ kV/cm). Only processes with primary components: nitrogen, oxygen, water and ethanol were taken into account since other secondary products poorly affected the EEDF because their concentrations are relatively small.

The calculated EEDF is shown in Fig. 2. One can see that $f(\varepsilon)$ function has a form characteristic for the case when $N_2$ is a plasma-forming gas.

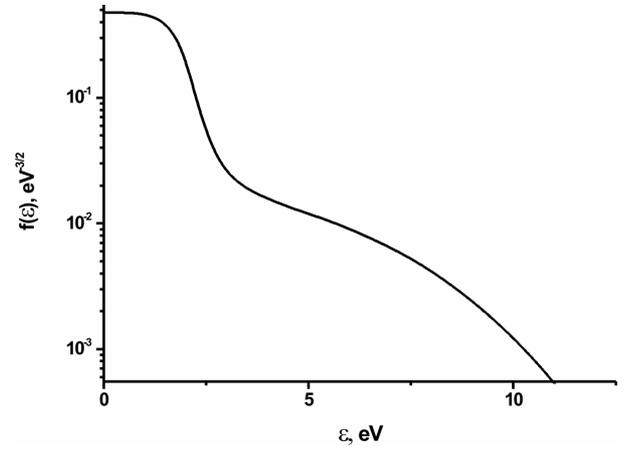

*Fig. 2. Calculated EEDF*

The gas components taking into account in plasma-chemical kinetics included primary components: $N_2$, $O_2$, $H_2O$ and $C_2H_5OH$, oxides of nitrogen and carbon, acids, and various hydrocarbons. The selected electron-molecular reactions for these components are presented in Table 1. The scheme of chemical reactions was compiled from [5, 6]. Besides, a water gas shift (WGS) reaction:

$$H_2O + CO \rightarrow CO_2 + H_2 \quad (\Delta H = -41 \text{ kJ/mol}) \quad (1)$$

was included in the scheme because of its importance [7] at the final stage of transformations outside the discharge.

*Table 1. Electron-molecular reactions taken into account in calculations of plasma kinetics in the mixture of air and ethanol and water vapors*

| N | Reaction | Reference |
|---|----------|-----------|
| 1 | $O_2 + e > O + O + e$ | [8] |
| 2 | $N_2 + e > N + N + e$ | [9] |
| 3 | $O_2 + e > O_2^+ + e + e$ | [10] |
| 4 | $N_2 + e > N_2^+ + e + e$ | [11] |
| 5 | $H_2O + e > OH + H + e$ | [10] |
| 6 | $O_2 + e > O + O(d) + e$ | * |
| 7 | $O_3 + e > O_2 + O + e$ | * |
| 8 | $N_2O + e > N_2 + O + e$ | * |
| 9 | $NO_2 + e > NO + O + e$ | * |
| 10 | $N_2O_4 + e > NO_2 + NO_2 + e$ | * |
| 11 | $N_2O_5 + e > NO_2 + NO_3 + e$ | * |
| 12 | $HO_2 + e > OH + O + e$ | * |
| 13 | $HO_2 + e > H + O_2 + e$ | * |
| 14 | $H_2O_2 + e > OH + OH + e$ | * |
| 15 | $OH + e > O + H + e$ | * |
| 16 | $N_2O + e > NO + N + e$ | * |
| 17 | $NO + e > N + O + e$ | * |
| 18 | $HNO + e > NO + H + e$ | * |
| 19 | $NO_3 + e > NO_2 + O + e$ | * |

| 20 | $HNO_2 + e > NO + OH + e$ | * |
|---|---|---|
| 21 | $HO_2NO_2 + e > NO_2 + HO_2 + e$ | * |
| 22 | $HNO_3 + e > OH + NO_2 + e$ | * |
| 23 | $HNO_3 + e > HO_2 + NO + e$ | * |
| 24 | $C_2 + e > C + C + e$ | [14] |
| 25 | $C_2H_2 + e > C_2H + H + e$ | [14] |
| 26 | $C_2H_3 + e > C_2H_2 + H + e$ | [14] |
| 27 | $C_2H_4 + e > H + C_2H_3 + e$ | [14] |
| 28 | $C_2H_5 + e > CH_2 + CH_3 + e$ | [14] |
| 29 | $C_2H_5OH + e > CH_3 + CH_2OH + e$ | [14] |
| 30 | $C_2H_5OH + e > C_2H_5 + OH + e$ | [14] |
| 31 | $C_2H_5OH + e > CH_3CHOH + H + e$ | [14] |
| 32 | $C_2H_6 + e > C_2H_5 + H + e$ | [14] |
| 33 | $C_2H_6 + e > CH_3 + CH_3 + e$ | [14] |
| 34 | $C_3H_6 + e > C_2H_3 + CH_3 + e$ | [14] |
| 35 | $C_3H_6 + e > C_3H_5 + H + e$ | [14] |
| 36 | $CH_2CO + e > CO + CH_2 + e$ | [14] |
| 37 | $CH_2CO + e > O + C_2H_2 + e$ | [14] |
| 38 | $CH_2O + e > CH_2 + O + e$ | [14] |
| 39 | $CH_2O + e > HCO + H + e$ | [14] |
| 40 | $CH_2O + e > CO + H_2 + e$ | [14] |
| 41 | $CH_2OH + e > CH_2 + OH + e$ | [14] |
| 42 | $CH_2OH + e > CH_2O + H + e$ | [14] |
| 43 | $CH_3CHO + e > CH_3 + HCO + e$ | [14] |
| 44 | $CH_3CHO + e > C_2H_4 + O + e$ | [14] |
| 45 | $CH_3CHO + e > CH_2HCO + H + e$ | [14] |
| 46 | $CH_3CHOH + e > C_2H_4 + OH + e$ | [14] |
| 47 | $CH_2HCO + e > CH_3 + CO + e$ | [14] |
| 48 | $CH_2HCO + e > C_2H_3 + O + e$ | [14] |
| 49 | $CH_2HCO + e > CH_2CO + H + e$ | [14] |
| 50 | $CH_3 + e > CH_2 + H + e$ | [14] |
| 51 | $CH_3O + e > CH_3 + O + e$ | [14] |
| 52 | $CH_3OH + e > CH_3 + OH + e$ | [14] |
| 53 | $CH_3OH + e > CH_2OH + H + e$ | [14] |
| 54 | $CH_3OH + e > CH_3O + H + e$ | [14] |
| 55 | $CH_4 + e > CH_3 + H + e$ | [12] |
| 56 | $CH_4 + e > CH_2 + H_2 + e$ | [13] |
| 57 | $CH + e > C + H + e$ | [14] |
| 58 | $CO_2 + e > CO + O + e$ | [12] |
| 59 | $CO + e > C + O + e$ | [12] |
| 60 | $HCO + e > CO + H + e$ | [14] |
| 61 | $HCOOH + e > HCO + OH + e$ | [14] |
| 62 | $C_2O + e > CO + C + e$ | [14] |
| 63 | $CH_2 + e > CH + H + e$ | [14] |
| 64 | $C_3H_4 + e > CH_3 + C_2H + e$ | [14] |
| 65 | $CH_3CH_2O + e > C_2H_5 + O + e$ | [14] |
| 66 | $H_2 + e > H + H + e$ | [14] |
| 67 | $C_2H + e > C_2 + H + e$ | [14] |
| 68 | $HCOH + e > CH_2 + O + e$ | [14] |
| 69 | $HCCO + e > H + C2H + e$ | [14] |
| 70 | $C_3H_5 + e > C_3H_4 + H + e$ | [14] |
| 71 | $CH_2CHO + e > CH_2CO + H + e$ | [14] |
| 72 | $CH_2CHO + e > C_2H_3 + O + e$ | [14] |
| 73 | $CH_3CH_2O + e > CH_3CHO + H + e$ | [14] |
| 74 | $CH_2CH_2OH + e > CH_2 + CH_2OH$ | [14] |
| 75 | $S-CH_2 + e > CH + H + e$ | [14] |
| 76 | $C_2H_4O + e > C_2H_4 + O + e$ | [14] |

### 3. Results and discussions

During the investigation an unexpected result was obtained at the calculation of the output hydrogen concentration dependence on the ethanol/water ratio: at equal amounts of ethanol and water in the solution the [$H_2$] output curve has the maximum (Fig. 3). This fact was confirmed in experiments.

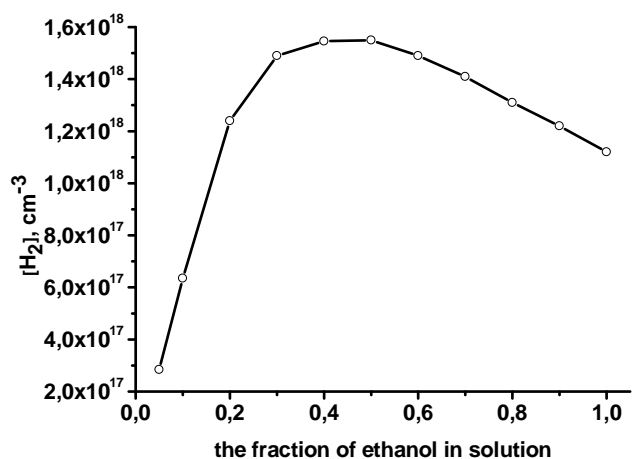

*Fig. 3. Dependence of output hydrogen concentration on the initial ethanol/water ratio*

The appearance of this maximum can be explained if consider the key reactions of the hydrogen generation. Among them the strongest effect gives the reaction of ethanol molecules with hydrogen atoms:

$$C_2H_5OH + H \rightarrow CH_3CH_2O + H_2 \qquad (2)$$

The main source of hydrogen atoms during the discharge is the fast e-impact dissociation of water molecules:

$$H_2O + e \rightarrow OH + H + e \qquad (3)$$

Therefore, the rate of the $H_2$ formation is proportional to the content of ethanol and water vapors. According to the model, the solution is ideal, and the concentrations of the specified components are determined by formulas:

$$[C_2H_5OH] = \frac{p_1}{kT} x \qquad (5),$$

$$[H_2O] = \frac{p_2}{kT}(1-x) \qquad (6),$$

where $x$ is a portion of ethanol in the solution, $p$ is the saturation vapor pressure at given temperature $T$. Thus, the $H_2$ yield is functionally quadratic on the ethanol content as $y \propto x(1-x)$ and it takes a maximum at $x=0.5$.

Since the exact value of gas temperature in the discharge cavity is not known, the calculations were produced for two points: $T=355$ K as assumed for the boiling temperature in the 50% ethanol-water solution and $T=323$ K as measured by the thermocouple in the solution in the working reactor. The comparison of calculated results and experimental data is presented in Fig. 4. One can see a rather good agreement for the main output syngas components, $H_2$ and CO at $T=323$ K.

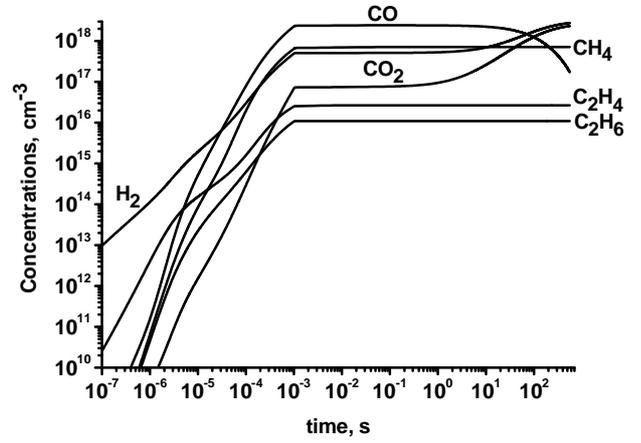

*Fig. 5. Time dependences of concentrations of the main syngas components*

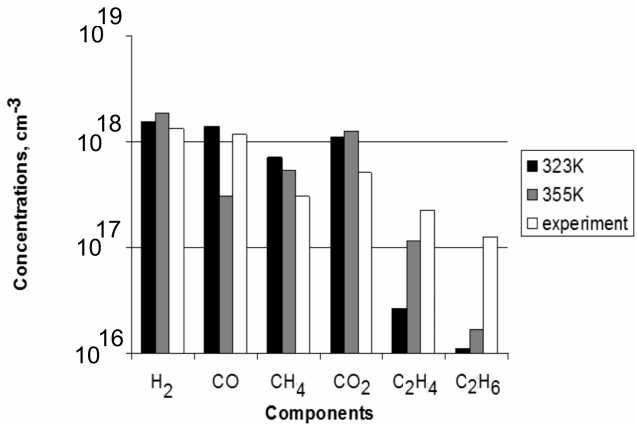

*Fig. 4. Comparison of calculated and experimental data*

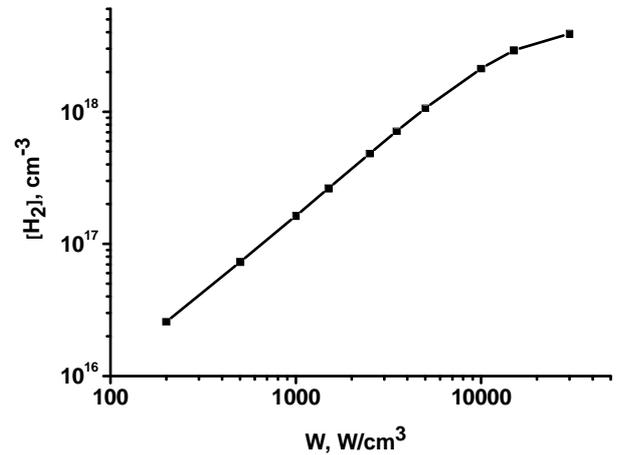

*Fig. 6. Dependence of output hydrogen concentration on the specific dscharge power*

The dynamics of variation of concentrations for some gas-phase components is illustrated in Fig. 5. One can see that during the discharge the production of $H_2$, CO and other species grows with the residence time up to $\sim 10^{-3}$s. Outside the discharge, after the time of $\sim 10$ s, the $H_2$ and CO have some changes due to the WGS reaction while $CH_4$ and other components at the final stage remain nearly constant as is seen in Fig. 5.

The dependence of the $H_2$ output on the specific discharge power $W/V$ is indicated in Fig. 6. At low values $W/V < 10^4$ W/cm$^3$ it is evidently approximated by a linear function. Such behavior is related to the fact that the $H_2$ generation is determined by the reaction of $H_2$ production from $C_2H_5OH$ (2) which rate depends on the number of H atoms generated mainly via the reaction of e-impact dissociation of $H_2O$ (3) which rate is directly proportional to the deposited discharge power. Consequently, the $H_2$ yield is also linearly increased with the parameter $W/V$. At high values $W/V > 10^4$ W/cm$^3$ the $H_2$ formation has influence of the direct e-impact dissociation of hydrogen molecules

$$H_2 + e \rightarrow H + H + e \qquad (7)$$

which leads to the $[H_2]$ output curve bending in Fig. 6.

### 4. Conclusions

In this work a new method of the plasma-assisted hydrogen production in the electric discharge in the mixture of air and water and ethanol vapors was studied theoretically and experimentally.

For the calculation of plasma kinetics the simplest model of the discharge burning was applied, which allows numerical simulations in agreement with experiments.

It was found that the maximal output of hydrogen is achieved in the case of equal amounts of ethanol and water in the solution. This was confirmed experimentally. It was also shown that the hydrogen output increased linearly with the specific discharge power and reached the saturation at high values.

### 5. References

1. V.V. Yukhymenko, V.Ya. Chernyak, V.V. Naumov, Iu.P. Veremii, V.A. Zrazhevskij. Combustion of ethanol-air mixture supported by transverse arc plasma // *Problems Atomic Sci. Technol., Series: Plasma Physics*. 2007, vol. 13, #1, p. 142-144.
2. V.Ya. Chernyak, S.V.Olszewski, V.V.Yukhymenko, E.V. Solomenko, I.V.Prysiazhnevych, V.V. Naumov, D.S. Levko, A.I. Schedrin, A.V. Ryabtsev, V.P.

Demchina, V.S. Kudryavtsev, E.V. Martysh, M.A. Verovchuk. Plasma-assisted reforming of ethanol in dynamic plasma-liquid system: experiments and modeling // *IEEE Trans. Plasma Sci. Special Issue on Plasma-Assisted Combustion.* 2008, vol.36 (in press).
3. A.G. Kalyuzhnaya, D.S. Levko, A.I. Schedrin. Comparison of varions methods of calculation of plasma kinetics in barrier discharge // *J. Tech. Physics.* 2008, vol. 78, #6, p. 122-126.
4. P.M. Golovinskii, A.I. Schedrin. // *J. Tech. Physics.* 1989, vol. 59, # 2, p. 51-56.
5. http://maeweb.ucsd.edu/~combustion/cermech/
6. N.M. Marinov. A detailed chemical kinetic model for high temperature ethanol oxidation // *Int. J. Chem. Kinet.* 1999, vol. 31, # 3, p. 183-220.
7. E.E. Shpilrain, S.P. Malyshenko, and G.G. Kuleshov. *Introduction to Hydrogen Energy.* M.: "Energoatom izdat", 1984, 264 p.
8. Yu.P. Raizer. *Physics of Gas Discharge.* Moscow: "Nauka", 1987, 592 p.
9. Y.S. Mok, S.W. Ham, I. Nam. Mathematical analysis of positive pulsed corona discharge process employed for removal of nitrogen oxides // *IEEE Trans. Plasma Sci.* 1998. vol. 26, # 5, p. 1566-1574.
10. H.C. Straub, P. Renault, B.G. Lindsay, K.A. Smith, R.F. Stebbings. Absolute partial cross sections for electron-impact ionization of $H_2$, $N_2$, and $O_2$ from threshold to 1000 eV // *Phys. Rev. A.* 1996. vol. 54, # 3, p. 2146-2153.
11. R. Rejoub, C.D. Morton, B.G. Lindsay, R.F. Stebbings. Electron-impact ionization of the simple alcohols // *J. Chem. Phys.* vol. 118, # 4, p. 1756-1760.
12. T. Shirai, T. Tabata, H. Tawara. Analytic cross-sections for electron collisions with CO, $CO_2$, and $H_2O$ relevant to edge plasma impurities // *At. Data Nucl. Data Tables.* 2001. vol. 79, # 1, p. 143–184.
13. D.A. Erwin, J.A. Kunc. Electron-impact dissociation of the methane molecule into neutral fragments // *Phys. Rev. A.* 2005, vol. 72, 052719-1-6.
14. In calculations of dissociation rates we used the dissociation cross section for oxygen biased by double value of threshold energy of the process.